\documentstyle[psfig,epsfig,prl,aps,floats]{revtex}
\begin{document}

\draft

\twocolumn[\hsize\textwidth\columnwidth\hsize\csname @twocolumnfalse\endcsname

\title{Island nucleation in thin-film epitaxy:  A first-principles
investigation}
\author{Kristen A.\ Fichthorn$^{1}$ and Matthias Scheffler$^{2}$}
\address{$^{1}$Department of Chemical Engineering,
The Pennsylvania State University, University Park, PA 16802, USA\\
$^{2}$Fritz-Haber-Institut der Max-Planck-Gesellschaft,
Faradayweg 4-6, D-14195 Berlin, Germany}
\maketitle

\begin{abstract}
We describe a theoretical study of the role of adsorbate interactions in island
nucleation and growth, using Ag/Pt(111) heteroepitaxy as an example. From
density-functional theory, we obtain the substrate-mediated Ag adatom pair
interaction and we find that, past the short range, a repulsive ring is formed
about the adatoms.  The magnitude of the repulsion is comparable to the
diffusion
barrier.  In kinetic Monte Carlo simulations, we find that the repulsive
interactions lead to island densities over an order of magnitude larger than
those predicted by nucleation theory and thus identify a severe limitation
of its
applicability. {\em Copyright 2000 by The American Physical Society.}

\end{abstract}

]

Island nucleation is often the first step in thin-film epitaxy and
is, thus, relevant to the synthesis of a wide variety of interfacial
materials. Achieving a quantitative understanding of the island
morphologies ({\em i.e.}, sizes, shapes, density, spatial distribution,
etc.) that develop in the initial stages of thin-film growth is also
important for fundamental reasons.  Since thin-film epitaxy frequently
occurs away from equilibrium, the kinetics of deposition and surface
diffusion play a key role in governing island morphology and there is great
variety in the resulting structures. Considering
shapes \cite{michely,brune1,ruggerone97,bogicevic1,ovesson99,stroscio,zhang97,guenther},
for example, islands can be fractal-like or compact and triangular, hexagonal,
square, rectangular, etc.. Each of these structures is a signature of an
intricate kinetic balance and reflects a complex set of interatomic
interactions
that is unique for each material.

Despite the complexity and potentially enormous variety in growth
morphologies, certain aspects of island nucleation and growth appear
to be common to many different systems. In a general description,
gas-phase species are deposited onto an initially bare solid
substrate with a rate $F$.  These species hop on the surface with
a rate $D =  \nu_{0}e^{-E_{b}^{0}/k_{B}T}$, where
 $\nu_{0}$ is
 the preexponential factor, $E_{b}^{0}$ is the diffusion-energy barrier for
an isolated species, $k_{B}$ is Boltzmann's constant, and $T$ is
temperature.   Hopping mediates the aggregation of adspecies into nuclei,
which either dissociate
with an energy barrier $E_{d,i}$, if they are below a critical size $i$,
or grow subsequently to become stable islands.  Initially, the
formation of island nuclei is the main process taking place. As the
surface coverage increases, it becomes increasingly likely that
deposited species will add to stable islands and promote their growth
instead of forming new nuclei. These general features can be captured
in a mean-field theory for the stable island density
$N_{x}$ \cite{ven73}.
In the island growth regime, this expression has the form
\begin{equation}
N_{x} \sim (F/D)^{i/(i+2)} \exp(-E_{d,i}/k_{B}T)^{1/(i+2)} .
\label{eq-island-density}
\end{equation}

Although the utility of a general expression cannot be overstated,
Eq.\ \ref{eq-island-density} cannot describe all aspects of thin-film
epitaxy and it is important to understand its limitations.  In the interest of
achieving a complete and predictive model for thin-film morphology, it is
clearly desirable to have an approach that is as free as possible from
arbitrary
parameters or assumptions. In this Letter, with an aim toward this ideal
approach, we present the results of a combined kinetic Monte Carlo (kMC) and
first-principles, density-functional theory (DFT) study of island
nucleation in a
model for the growth of Ag on a monolayer (ML) of Ag on Pt(111). Our choice of
this model system was motivated by intriguing results from recent,
low-temperature, scanning-tunneling microscopy (STM)
studies \cite{fischer99,barth,brune95}, in which
Eq.\ \ref{eq-island-density} was
used to obtain the energy barrier and preexponential factor for adatom
hopping.
Shown in Table I are the parameters obtained in these studies.

\begin{table}[b]
 \begin{tabular}{c|c|c|c}
 System & $E_{b}^{0}$ (meV)    & $\nu_{0}$ (s$^{-1})$   & Ref.  \\
\hline
Al on Au(111) &  30 & $7 \times 10^{3}$ & \cite{fischer99}\\
\hline
 Al on Al(111) &  42 & $8 \times 10^{6}$ &  \cite{barth} \\
\hline
 Ag on 1-ML-Ag/Pt(111) &  60  & $10^{9}$ &   \cite{brune95} \\
 \end{tabular}
 \medskip
 \caption{Experimentally determined diffusion-energy barriers and
preexponential factors.}
 \vspace{-1ex}
\end{table}

A striking feature of the experimental results is that the preexponential
factors are significantly smaller than would be anticipated for systems
such as these.  For example, from {\em ab initio} calculations, Ratsch and
Scheffler \cite{ratsch98} find a preexponential factor of $\nu_{0} = 1.3
\times 10^{12}$ s$^{-1}$ for a Ag adatom on 1-ML-Ag/Pt(111), with a
diffusion barrier of $E_{b}^{0}$ = 63 meV.  Inserting the experimental and
theoretical values for the diffusion parameters into
Eq.\ \ref{eq-island-density} in the low-temperature limit where $i = 1$ and
$E_{d,i} = 0$, we see that the experimental island densities are about an order
of magnitude higher than predictions based on the theoretical diffusion
parameters.  Here, we investigate the origins of this discrepancy.  Our DFT-kMC
model includes many features of the complex potential-energy surface
experienced
by Ag adatoms during thin-film growth and is free from several of the
assumptions
in Eq.\ \ref{eq-island-density}.  We find that one of these assumptions -- that
interactions between adsorbed species do not extend beyond a short range -- is
violated. For systems with low diffusion-energy barriers [such as Ag on
1-ML-Ag/Pt(111)], we show that these long-range, adatom-adatom interactions
play
an important and previously underestimated role in island nucleation and
growth.

 The DFT calculations \cite{bockstedte97} are performed using the
 plane-wave, pseudopotential  \cite{fuchs99} method within
 the generalized gradient approximation  \cite{perdew96}.
 Previously, Ratsch {\em et al.} \cite{ratsch97} showed in DFT
 calculations that the diffusion-energy barrier of an Ag atom on
 the 1-ML-Ag/Pt(111) substrate is essentially the same as that on a
strained Ag(111) substrate, in which Ag is given the lattice constant of
Pt.  Thus, to model the heteroepitaxial system we use strained Ag(111), in
which
the lattice constant is set to a value of 4.01 \AA.  This value is 4.61 percent
smaller than our calculated lattice constant for bulk Ag. We use the supercell
approach to describe the surface, which is modeled as a $(4\times 4\times 4)$
slab with a vacuum spacing of five interlayer distances. The cut-off energy
is 50
Ry and we use 4 ${\bf {k}}$ points to sample the full surface Brillouin zone.
The top layer of a bare slab is fully relaxed.  Subsequently, an adatom is
placed in a binding site (fcc and hcp three-fold hollow sites), and its
height is
optimized with respect to the fixed substrate.  To calculate adatom interaction
energies, two (or more) adatoms are placed on the relaxed (and fixed) substrate
with heights fixed to values from the single-adatom calculations.  In this
way, we
seek to isolate the electronic interaction between adatoms in binding
sites. With
simultaneous relaxation of both the adatoms and surface atoms, we can resolve
the role of substrate-mediated, elastic interactions in the total interaction
energy.  Full relaxation of a few trial structures and inspection of the forces
in our partially relaxed slabs indicates that elastic interactions are not
highly dependent on adsorbate configuration and that our results will
change by 10
meV or less with full relaxation.

 The total interaction energy $\Delta E$ for a periodic slab
 containing $N$ adatoms, of which $M$ are at binding site $a$ and $(N-M)$
are at
binding site $b$, is given by
$\Delta E = E_{S+N}^{a,b} - ME_{S+1}^{a} - (N-M)E_{S+1}^{b} + (N-1)E_{S}$.
 Here, $E_{S+N}^{a,b}$ is the total energy of a slab with $N$ adatoms,
$E_{S+1}^{a}$ and $E_{S+1}^{b}$ are the total energies
 of slabs containing one adatom, and $E_{S}$ is the total energy of a bare
slab.  In the DFT supercell approach, the total interaction
energy is comprised of interactions between different adatoms in the slab
and interactions between adatoms in the slab and the periodic-image
adatoms.  We
can express $\Delta E$ as a function of these interactions using the
lattice-gas
Hamiltonian approach (see, {\em e.g.}\cite{stampfl99}), which yields
 \begin{eqnarray}
 \Delta E &=&
 \frac{1}{2}\sum_{i,j}V^{(2)}({\bf{R}}_{i,j}) n_{i}n_{j} \nonumber \\
 &&+\frac{1}{3} \sum_{i,j,k}V^{(3)}({\bf {R}}_{i,j},
 {\bf{R}}_{i,k}) n_{i}n_{j}n_{k} + \ldots
 \label{eq-lattice-gas}
 \end{eqnarray}
 Here, the summations run over all sites $i$ in the slab and all
 sites $j$ and $k$ in the supercell (which includes both the slab and its
periodic images),
$n_{m}$ is unity if site
$m$
 ($m = i,j,k$) is occupied and zero, otherwise, $V^{(2)}({\bf R}_{i,j})$ is
the pair interaction between two adatoms on sites $i$ and $j$, and
$V^{(3)}({\bf R}_{i,j}, {\bf R}_{i,k})$ is the
 trio interaction between three adatoms on sites $i$, $j$, and $k$.  We
neglect higher-order interactions.
Another assumption implicit in Eq.\ \ref{eq-lattice-gas} is that the
interaction between adatoms at a fixed distance is independent of
whether these atoms occupy fcc or hcp sites. We confirmed this
assumption in one trial calculation.  Finally, the adatom binding energies
on fcc and hcp sites are virtually equal: The fcc site is favored by less
than 3
meV.

Thus, for a given adatom configuration, we express $\Delta E$ as a sum of pair
and trio interactions with unknown
coefficients.  From 18 different configurations, we obtain a system of linear
equations and solve these for pair-interaction coefficients
up to the $13^{\rm th}$-neighbor, as well as for 5 different trio
interactions.  We assume that all other interaction
coefficients are zero.  To verify our parameterization of Eq. (2), we used our
interaction parameters to predict the total interaction energy in several,
additional test structures.  All of the predicted values agreed well with
values
from DFT calculations.

\begin{figure}[t]
 \psfig{figure=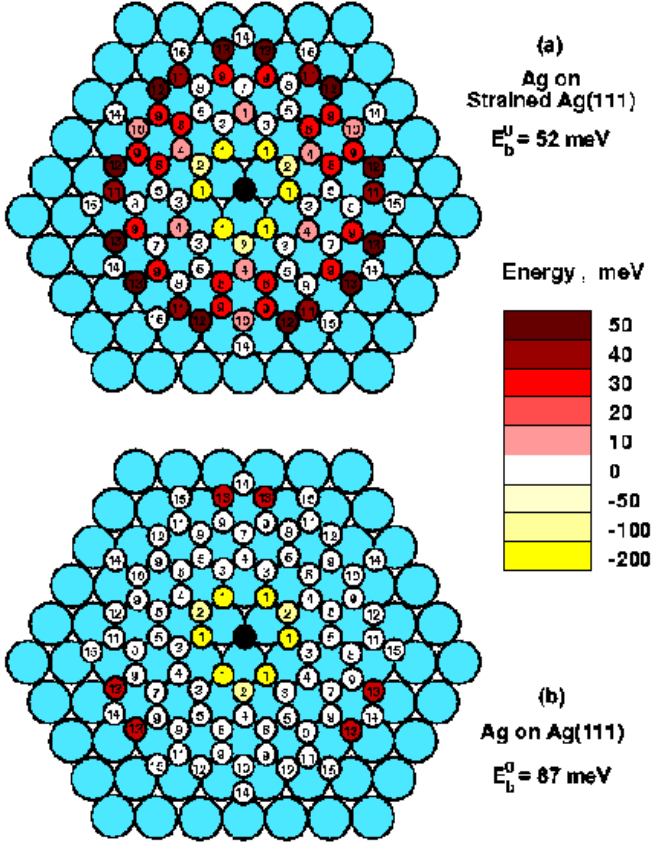,width=\columnwidth}
 \caption{Pair interaction energy as a function of distance from a central
 adatom, shown in black, for (a) Ag on strained Ag(111) and (b) Ag on Ag(111).
Also noted are the diffusion-energy barriers, $E_{b}^{0}$, of isolated
adatoms on these substrates.}
 \label{fig-pair-potential}
\end{figure}

The pair interaction is shown in Fig.\ \ref{fig-pair-potential}, where we
also show results for Ag on {\em unstrained} Ag(111).  For both surfaces, this
interaction is strongly attractive at the nearest-neighbor distance and
repulsive
at longer distances.  It is interesting to consider the origins of the
long-range
repulsion. At these distances, the interaction could be due to
substrate-mediated
elastic interactions or of electronic origin \cite{einstein}.  Since we
find that
elastic interactions play a small role here, the repulsion is primarily an
electronic effect.  Each adatom induces a small perturbation in the electron
density, which decays with distance from the adatom in an oscillatory manner.
The asymptotic tail, which is expected to decay with distance $d$ as
$d^{-5}$ (or
as $d^{-2}$, if a partially filled surface state is involved), is a
Friedel-type oscillation. Friedel oscillations have been imaged as concentric,
ring-like, features around defects in low-temperature STM studies of several
noble-metal surfaces [including Ag(111)] \cite{briner}.  We expect the Friedel
tail to extend to much longer distances than can be probed in DFT
calculations.
However, interactions associated with the Friedel tail should be weaker
than those
probed here.  Thus, the central-ring interaction resolved here will have
the most
significant ramifications for thin-film morphology.

 %
 %

From Fig.\ \ref{fig-pair-potential}(a), we see that the
magnitude of the repulsive ring for Ag on strained Ag(111)
is comparable to the diffusion-energy barrier for an isolated
adatom.  For Ag on Ag(111), it appears that the repulsive interaction is weaker
and the diffusion-energy barrier is larger.  The diffusion barriers
reported here
are obtained with full relaxation of both the first-layer slab atoms and the
adatom.  Our barriers are in good agreement with experimental
values \cite{brune95} for Ag on 1-ML-Ag/Pt(111) (60 meV) and on Ag(111) (97
meV)
and with those of Ratsch and Scheffler \cite{ratsch98}.  If the interaction
energy and the diffusion barrier are of comparable size, we expect interatomic
interactions to significantly influence adatom diffusion and island formation.
Since Eq.\ \ref{eq-island-density} neglects the influence of long-range
interactions, it is unclear if this expression is accurate under these
circumstances.

To resolve the effect of long-range interactions on thin-film growth, we
developed
a kMC model employing the general method of Fichthorn and
Weinberg \cite{fichthorn91} and incorporating the pair potential for
Ag on strained Ag(111) shown in
 Fig.\ \ref{fig-pair-potential}(a).  In the initial stages of thin-film epitaxy,
the surface coverage is low and pair interactions are likely to be the only
significant interactions governing island nucleation and growth\cite{comment1}.
In our kMC model, atoms are deposited onto a
fcc(111) substrate with a rate of $F = 0.1$ ML/s.  An adatom hops from site
$i$ to site $j$ with a rate given by $D_{i\rightarrow j} =  \nu_{0}e^{-E_{i
\rightarrow j}/k_{B}T}$, where $E_{i \rightarrow j}$ is the energy barrier
to hop from site $i$ to $j$.
For the hopping-rate parameters, we use $\nu_{0} = 10^{12}$
s$^{-1}$ \cite{ratsch98}.  The energy barrier is given
 by $E_{i \rightarrow j} = E^{\ddagger}_{i,j} - E_{i}$, where $E_{i}$ is the
energy with an atom at site $i$ and $E^{\ddagger}_{i,j}$ is the energy of the
transition state between sites $i$ and $j$.
 In general, $E^{\ddagger}_{i,j}$ should depend on both $E_{i}$ and $E_{j}$.
Considering possible permutations of adatom configurations with $13^{\rm
th}$-neighbor interactions, $\sim 10^{14}$ different, diffusion-energy barriers
could occur.  To make the problem tractable, we adopt a simple model, in which
$E_{i \rightarrow j} = E_{b}^{0} + \frac {1}{2}(E_{j} - E_{i})$.
All of the quantities in this equation are
obtained from DFT calculations.  We
 have tested this equation for a trial geometry in which an adatom
with four fcc $9^{\rm th}$ neighbors hops to a nearest-neighbor hcp site
where it
has two $7^{\rm th}$ and two $12^{\rm th}$ neighbors.  From our simple
model, we find $E_{i \rightarrow j}$ = 53 meV, which is in
remarkable agreement with the value of 46 meV from DFT calculations.

\begin{figure}[t]
 \epsfig{figure=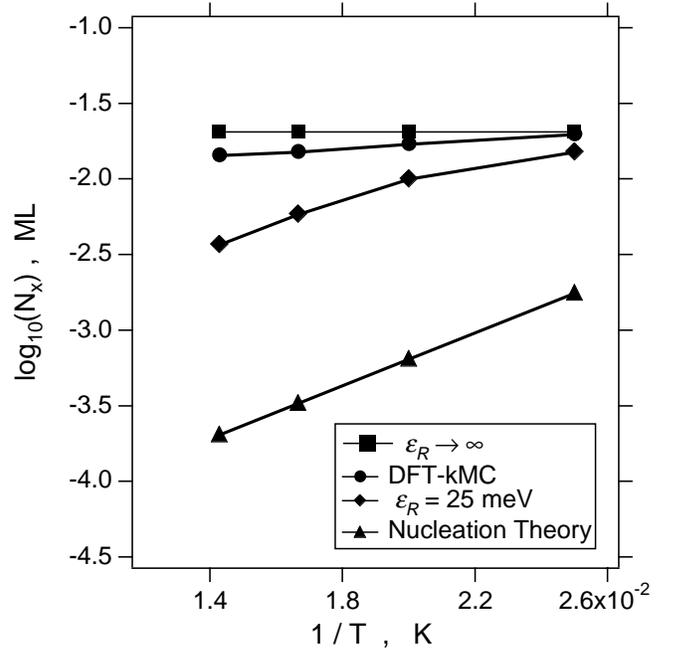,width=\columnwidth}
 \caption{Arrhenius plot of the island density as a function of temperature
from an impermeable repulsive ring (squares), the DFT-kMC model
(circles), repulsive ring with $\varepsilon_{R}$, = 25 meV (diamonds), and
nucleation theory (triangles).}
 \label{fig-kMC}
 \end{figure}

We simulated thin-film epitaxy over
 temperatures ranging from 40-70 K and determined island densities in the
beginning of the island growth regime.  These low temperatures are in the
range of the experimental studies (cf., Table I). At such low
temperatures, Eq.\ \ref{eq-island-density} reduces to the form $N_{x} \sim
(F/D)^{1/3}$. Fig.\ \ref{fig-kMC} shows an Arrhenius plot of the island
density
from our DFT-kMC model as a function of temperature.  Also shown in
Fig.\ \ref{fig-kMC} is
the island density predicted by nucleation theory for the values of $F$,
$\nu_{0}$, and
$E_{b}^{0}$
 used here.  To
quantitatively compare nucleation theory with the simulations, a
proportionality
coefficient
$\eta$ is needed in Eq.\ \ref{eq-island-density} ({\em i.e.}, $N_{x} =
\eta(F/D)^{1/3}$).  This coefficient is related to the efficiency of the
islands in capturing adatoms.  Using a self-consistent approach, $\eta$ =
0.25 \cite{bales} and values of $\eta$ ranging from 0.2 to 0.23 have been
found in kMC simulations of Ag island nucleation on Pt(111) \cite{brune99}.
Here, we use $\eta$ = 0.25.

 %
 %


In Fig.\ 2, we see that the DFT-kMC island densities are an order of magnitude
(or more)
 above the theoretical values.  To understand this, we construct
a caricature
 model, in which we replace the set of pair interactions shown in
 Fig.\ \ref{fig-pair-potential}(a)
 with a nearest-neighbor attractive interaction and a uniform, repulsive
 ring of strength $\varepsilon_{R}$ at distances 10-13.  By varying the
 magnitude of $\varepsilon_{R}$, we span the entire range of possible behaviors
in this system.
 As $\varepsilon_{R} \rightarrow \infty$, the island density assumes a
 constant, maximum value that is
 independent of temperature (cf.,
Fig.\ \ref{fig-kMC}).  This is
 because
 island nucleation can only occur when one atom is deposited within the
 repulsive ring of another and it is governed by the temperature-independent
deposition rate.  In this regime, many adatoms are isolated by repulsion
in the initial stages of deposition.  Each isolated adatom becomes a stable
island when another atom is deposited into its ring and the resulting island
density is significantly higher than in the absence of such a ring.

 As $\varepsilon_{R}$ is decreased, diffusing adatoms are
 increasingly able to surmount the ring barrier and a second channel for island
nucleation and growth (via
 long-range, adatom diffusion) opens up.  The extent to
 which long-range diffusion contributes to island nucleation and growth
 depends on the temperature.  In Fig.\ \ref{fig-kMC}, we see that at 40 K, the
DFT-kMC island density is the same as that for an
 infinitely repulsive ring ({\em i.e.}, diffusing adatoms are unable to
 penetrate the
 ring on the time scale for nucleation). As the temperature increases, adatoms
 are increasingly able to penetrate the ring barrier to aggregate and add
to existing islands via
 long-range diffusion.
 Consequently, the island density decreases with increasing temperature.
 It is interesting to note that for the conditions studied, even a
 relatively weak repulsive ring with $\varepsilon_{R}$ = 25 meV can lead to
 significantly higher island densities than those predicted by nucleation
 theory.

Returning to our discussion of the experimental results shown in Table I, we
point out that the order-of-magnitude difference between the island densities
predicted from {\em ab initio} calculations \cite{ratsch98} and those found
experimentally for Ag on 1-ML-Ag/Pt(111) \cite{brune95} is also seen in our
study, comparing the island densities predicted by nucleation theory to those
found in our DFT-kMC ``computer experiments'' (cf., Fig.\ \ref{fig-kMC}).
Thus, we
conclude that our results can explain the theoretical-experimental gap in the
island density for Ag on 1-ML-Ag/Pt(111).  Further, our results indicate
that for
Ag on Ag(111), the theoretical-experimental gap should be weaker or
non-existent.  This result is also consistent with a comparison of theoretical
diffusion parameters for Ag on Ag(111) \cite{ratsch98,ratsch97}
($\nu_{0}=8.2 \times 10^{11}
$s$^{-1}$, $E_{b}^{0}= 82$ meV) to experimental values \cite{brune95} obtained
using Eq.\ \ref{eq-island-density} ($\nu_{0}= 2 \times 10^{11} $s$^{-1}$,
$E_{b}^{0}=$ 97 meV).  Finally, Bogicevic and co-workers\cite{competition}
recently found similar DFT and kMC results for both Al(111) and Cu(111)
homoepitaxy.

 Thus, we conclude that long-range, electronic, substrate-mediated adatom
 interactions exist and, if their strength is comparable
 to the diffusion barrier, they can significantly
 influence surface diffusion and the growth morphology in thin-film
epitaxy.  For
Ag on strained Ag(111), the adatom pair interaction becomes repulsive past the
short range and the repulsion forms a ring around isolated adatoms.  The
 magnitude of the repulsion is comparable to the
 diffusion barrier.  By inhibiting island nucleation
 and growth via long-range adatom diffusion, these interactions lead to
 island densities that are substantially larger than those
 predicted by nucleation theory.

 We acknowledge helpful conversations with A. Bogicevic, H. Brune, P.
Kratzer, C.
Ratsch,
 A. Seitsonen, and C. Stampfl.  Support for this work is from the Alexander
 von Humboldt Foundation and the NSF (DMR-9617122).



\begin{references}
\bibitem{comment1} We find that trio interactions are attractive and
significant at high adatom densities.  They are unimportant at the medium and
longer distances probed here and we will discuss them elsewhere.
\end{references}
 \end{document}